# Selecting between two transition states by which water oxidation intermediates on an oxide surface decay

Xihan Chen[1,2], Daniel J. Aschaffenburg[1], and Tanja Cuk[1, 2, 3, 4*]

[1] Department of Chemistry, University of California, Berkeley, Berkeley, CA.

[2] Chemistry and Nanoscience Center, National Renewable Energy Laboratory, Golden, CO.

[3] Chemical Sciences Division, Lawrence Berkeley National Laboratory, Berkeley, CA.

[4] Department of Chemistry and Renewable and Sustainable Energy Institute (RASEI), University of Colorado, Boulder, CO.

While catalytic mechanisms on electrode surfaces have been proposed for decades, the pathways by which the product's chemical bonds evolve from the initial charge-trapping intermediates have not been resolved in time. Here, we discover a reactive population of charge-trapping intermediates with states in the middle of a semiconductor's band-gap to reveal the dynamics of two parallel transition state pathways for their decay. Upon photo-triggering the water oxidation reaction from the n-SrTiO$_3$ surface with band-gap, pulsed excitation, the intermediates' microsecond decay reflects transition state theory (TST) through: (1) two distinct and reaction dependent (pH, T, Ionic Strength, and H/D exchange) time constants, (2) a primary kinetic salt effect on each activation barrier and an H/D kinetic isotope effect on one, and (3) realistic activation barrier heights (~0.4-0.5 eV) and TST pre-factors (~$10^{11}$ -$10^{12}$ Hz). A photoluminescence from mid-gap states in n-SrTiO$_3$ reveals the reaction dependent decay; the same spectrum was previously assigned by us to hole-trapping at parallel Ti-O$^\bullet$-Ti (bridge) and perpendicular Ti-O$^\bullet$ (oxyl) O-sites using *in-situ* ultrafast vibrational and optical spectroscopy. Therefore, the two transition states are naturally associated with the decay of these respective intermediates. Furthermore, we show that reaction conditions select between the two pathways, one of which reflects a labile intermediate facing the electrolyte (the oxyl) and the other a lattice oxygen (the bridge). Altogether, we experimentally isolate an important activation barrier nfor **water oxidation, which is necessary for designing water oxidation catalysts with high O₂ turn-over. Moreover, in isolating it, we identify competing mechanisms for O₂ evolution at surfaces and show how to use reaction conditions to select between them.**

Electrochemical reactions are important for storing electrical into chemical energy and the most complex, multi-step reactions are thought to occur at electrode surfaces[1,2]. The surface plays the role of a catalyst that reconfigures reactants into products through a series of electron transfer and chemical bond-making and breaking events[2,3]. The mechanistic pathways by which these events occur differentiate catalysts and have been proposed for decades by electrochemical investigations that track product evolution[4-6]. However, significant questions remain un-resolved, such as how to select between faster and slower pathways[7,8], and the nature of transition states related to critical reaction steps[5,9], including the formation of the product's chemical bonds[10,11]. More generally, the properties of electrochemically controlled catalysis at an electrode surface—such as how charge is efficiently funneled to catalytic events and the existence of multiple and tunable pathways—have yet to be explained by the dynamics that connects the intermediate steps to product evolution.

The multi-electron transfer water oxidation is a canonical, electrochemical reaction that can occur efficiently both at an electrode[12] and in homogenous form[13] and for which there have been a number of mechanistic studies on the intermediate steps. In Photosystem II, the oxygen evolving complex (OEC)[14] that evolves O₂ receives charge from the photo-induced charge separation occurring in the chromophore antenna[15], and the configuration of the OEC after each charge transfer has been investigated by mid-infrared[16], optical[17], and x-ray studies[14,15,18]. Similar architectures for solar-to-fuel generation as Photosystem II are proposed with dye-molecules injecting charge into delocalized electronic states in the material, from which charge trapping creates local active regions on the surface that, ideally, mimic the configuration of the OEC[19].

The one-electron charge-trapping intermediates expected to initiate O-O bond formation have been directly investigated at transition metal oxide surfaces to reveal forms such as Ti-O•, Fe(IV)=O, and Co(IV)=O.[20-22] We note that in the time-resolved mechanistic studies, a photo-trigger is utilized, which means that the potential of the initially created holes is limited to the optically excited electronic levels within the dye-molecule or semiconductor. On the other hand, the subsequent steps of catalysis—once charge is separated from the light pulse—should be at the very least analogous to electrochemical systems run at the potential of the surface trapped charge, where intermediates such as Ti-O•, Fe(IV)=O, Co(IV)=O are also anticipated. In contrast to homogeneous catalysts such as the OEC, where only one or two should occur within a cycle, the one-electron intermediates of the reaction at a surface are expected to significantly populate it, which can reflect mechanisms involving many inter-connected and spatially separated sites[23,24].

Thus far, however, the dynamics of how the charge trapping intermediates decay into the next event in the cycle has not been revealed through distinct activation barriers. The difficulty has been in isolating the intermediates and their time-evolution; either a product[5,6] or generic hole signature[25,26] has been followed, or charge-trapping intermediates without the necessary time resolution.[21,22] In previous work, we were the first to capture the picosecond population of the surface by the charge-trapping, one-electron intermediates of water oxidation through their electronic levels, which create states in the middle of the semiconductor band gap. At the n-SrTiO$_3$ surface, we identified the oxyl radical (Ti-O•) which terminates the surface and the bridge intermediate parallel (Ti-O•-Ti) to the surface, by a vibration assigned to the oxyl and their distinct optical polarization dipoles.[27] The vibration unique to a site-localized oxyl radical, a sub-surface motion of the O right below it, occurs with the same rise-time as the mid-gap levels[20]. Since these charge-trapping intermediates were captured at their "birth", and under conditions of a highly efficient photo-trigger (75% charge separation) of Faradaic O$_2$ evolution[28], we are uniquely in the position to track their decay through the next event within the water oxidation cycle.

In this manuscript, by probing the microsecond decay of the mid-gap band with a 400 nm optical probe in reflectance (Supplementary Figure 1) after photo-triggering Faradaic O$_2$ evolution from the n-SrTiO$_3$ (Nb 0.1%)/aqueous interface (Supplementary Figure 2), we reveal two distinct transition states for the intermediates' decay by varying pH, ionic strength, H/D exchange, and temperature. The transition states are defined by two consistent time constants over a large range of reaction conditions, reasonable activation barriers and pre-factors, a kinetic isotope effect upon H/D exchange on one barrier, and a primary kinetic salt effect on both. The primary kinetic salt effect defines two or more localized Ti-O• intermediates in each transition state. We follow the same mid-gap band as that assigned to the oxyl and bridge at ultrafast (picosecond) time-scales. Therefore, each decay pathway is naturally associated with a transition state that creates a new termination at a site previously occupied either by an oxyl or a bridge intermediate. Further, the two activation barriers occur in parallel with reaction conditions favoring either the labile (oxyl) or lattice (bridge) intermediate. The likely next event triggered by these activation barriers is the formation of the O-O bond, thought to be a rate-limiting step of the reaction. Recently, such multiple and tunable routes of $O_2$ evolution were underscored by the findings of "slow" and "fast" sites[22] on the millisecond time-scale and two mechanisms separable by the extent of surface excitation[25]. Here, we uniquely isolate the activation barriers and character of the transition states, show that they can be described by canonical TST theory, and demonstrate how to utilize reaction conditions to favor one pathway over the other.

**Results**

Figure 1 depicts the electronic and vibrational spectra of the one-electron charge-transfer intermediates on $SrTiO_3$. Depicted in Figure 1A, the spectra of the oxyl radical's vibration is a sub-surface motion of the O right underneath the surface/interfacial Ti-O•. Depicted in Figure 1B, the optical spectrum of oxyl and bridge intermediates results from emissive transitions between conduction band and their mid-gap electronic states. Both of these spectral signatures were defined at the ultrafast time-scales when the intermediates are first created. In particular, the optical spectrum has been uniquely assigned by us to holes trapped at O-sites: at ultrafast time-scales, this is the spectrum, decomposed from broad band white light, that arises with the same dynamics with which oxyl and bridge intermediates, probed electronically at 400 nm and vibrationally at 800 cm$^{-1}$ by their characteristic optical and mid-infrared transition dipoles, populate the surface.[20,27] Importantly, the electronic and vibrational spectra maintain their shape through the observed decay at microsecond time-scales, such that they should track the subsequent evolution of oxyl and bridge intermediates. A deviation at low wavelengths is seen at longer times, likely due to spectral distortions at the edge of the sapphire-generated white light spectra.[29] Notably, the optical spectrum at each time point tracks the characteristic photoluminescence spectrum from $SrTiO_3$[27], shown in Fig. 1B for the same electrolyte but at open circuit and under steady state excitation. We also note that both the electronic and vibrational spectra are similarly quenched by methanol at microseconds, as expected for the hole-trapping intermediates (Supplementary Figure 3).

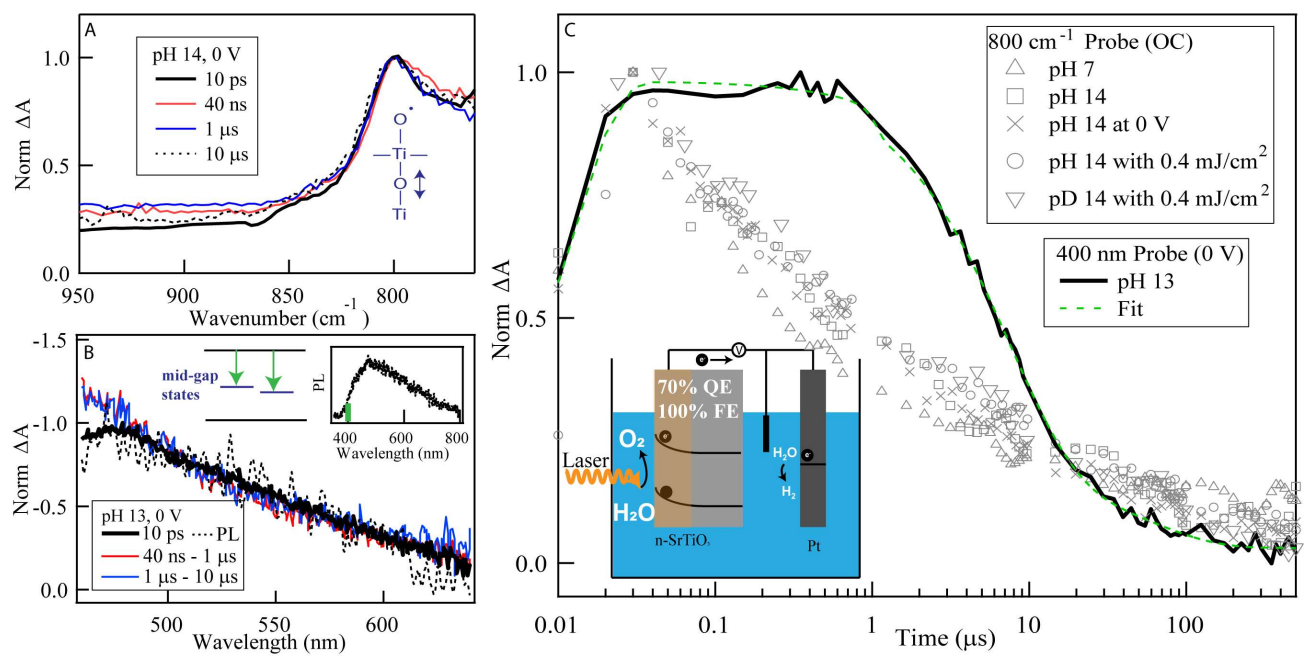

**Figure 1. Spectra and kinetics of the oxyl's sub-surface vibration and the mid-gap levels** (A) Time evolution of representative mid-IR spectrum (oxyl radical's sub-surface vibration) probed in p-polarization at 10 ps (black), 40 ns (red), 1 µs (blue), and 10 µs (black, dashed) at pH 14, 0 V vs. Ag/AgCl. (B) The spectrum of the optical transitions of the intermediates measured after excitation at 10 ps (black, solid line), 40 ns - 1 µs (time-averaged, red) and 1 µs - 10 µs (time-averaged, blue) compared with Photoluminescence spectra (black, dashed) at pH 13, 0 V vs Ag/AgCl. The inset of (B) shows the PL spectrum from 350 nm to 800 nm. The red vertical line indicates the optical probe wavelength. The cartoon shows the carriers relaxing from the conduction band to midgap state, resulting in PL emission. (C) Kinetics of transient response at the peak of the vibration (800 cm$^{-1}$) for varying experimental conditions (pH, H/D, OC, 0 V vs Ag/AgCl) and near the peak of the optical spectrum (400 nm, solid line) for pH 13 NaOH solution at 0 V vs Ag/AgCl. The decay of the optical spectrum is well-fit with a bi-exponential after a pulse limited rise (green dotted line). The inset depicts the photo-electrochemical cell. Under laser experimental condition, n-SrTiO$_3$ reached 70% quantum efficiency and 100% faradic efficiency towards oxygen evolution.

Figure 1C shows the microsecond decay of the optical (400 nm) and vibrational signatures (800 cm$^{-1}$) of the intermediates. As shown there, we find that the dynamics probed by the oxyl's vibration (at 800 cm$^{-1}$) are insensitive to reaction conditions, for: open (no current flow) or closed (current flow at 0 V vs Ag/AgCl) circuit, pH 7 through pH 14, H vs D, and surface excitation from 3% to 30%. The closed circuit conditions are taken at pH 14 due to reactant diffusion limitations within the attenuated total reflection geometry of the mid-infrared evanescent probe[20]. The vibrational transition remains p-polarized throughout, identifying a vibration consistently perpendicular to the sample plane (Supplementary Figure 4). Due to the insensitivity to reaction conditions, the time-trace represents oxyl radicals which have yet to react.

We now turn to how these mid-gap optical transitions exhibit a well-defined, bi-exponential decay highly dependent on reaction conditions. These will be probed with the narrowband, 400 nm pulse at the edge of the photoluminescence spectrum, which explicitly avoids overlapping transitions related to VB holes (> 650 nm)[30,31], Nb-dopants[32], and O-vacancies[33,34]. Importantly, 400 nm light also avoids optical transitions related to the O-O bond, which could occur at microsecond time-scales, but whose electronic states pattern molecular orbitals and should not lie in the middle of the gap[35,36]. Likely because it avoids these alternatives, the 400 nm wavelength cleanly mapped the rise time of populating the surface with the one-electron intermediates onto the vibrational (800 cm$^{-1}$) signature of the oxyl. While highly sensitive to reaction conditions, the bi-exponential decay is well reproduced for different trials and sample batches, with negligible scatter in comparison to the reaction-independent time-trace of the oxyl's vibration (Supplementary Figure 5, Supplementary Figure 6). Throughout, the applied potential is kept at 0 V vs Ag/AgCl, at the center of the potential-independent photo-current (Supplementary Figure 2). Although the applied potential does alter the VB hole potential within this photo-current[30], it does not modify the rates of formation of oxyl and bridge intermediates[27] or, as reported here, the rates of their decay[27] (Supplementary Figure 7). Therefore, the activation barrier ($\Delta G^*$) of the reactions we investigate is defined, on the reactant side, by the electronic potential energy of surface hole traps for Ti-O$^\bullet$ and Ti-O$^\bullet$-Ti intermediates. In the following, we report on how the bi-exponential decay rates depend on a full range of reaction conditions (pH, H/D exchange, and ionic strength); for all of these conditions, the quantum efficiency of charge separation is greater than 70% (Supplementary Figure 8, 9). We maintain the surface excitation at 3% (0.04 mJ/cm$^2$), where we have confirmed that separated holes lead to 100% Faradaic $O_2$ evolution[28]. The raw data are fit with: $\Delta A(t) = A_1 e^{-t/\tau_1} + A_2 e^{-t/\tau_2}$ (see General Fitting Function in SI for details). Since the exponential fits were carried out from the initial decay, ($A_1 + A_2$) initially represents the total population of oxyl and bridge intermediates observed at ultrafast timescales. As the decay evolves, however, each route ($A_1$ and $A_2$) is better described by the population of one-electron intermediates, which could include many different configurations of them, connected to a particular transition state. In the bi-exponential fit, since the time-traces are fit from the start of the decay, $\tau_1$ and $\tau_2$ represent two parallel routes rather than sequential events within the same cycle. We note that the data were also analyzed by several other fitting procedures for pH 7 to pH 13 at constant ionic strength (Supplementary Figure 10 – 13 and Supplementary Table 1 - 4). A single exponential has difficulty defining the latter part of the decay (> 10 µs). A bi-exponential fitting procedure in which $\tau_1$ and $\tau_2$ represent sequential rather than parallel events also has difficulty with the latter part of the decay, and furthermore, returns a standard deviation on the second time constant much larger than its value. A triple exponential over-fit the data, with a very large standard deviation on the slowest and third time constant. While all of these alternative fitting procedures give poorer fits, the bi-exponential fit of two parallel routes is furthermore the only one for which $\tau_1$ and $\tau_2$ are constant over a significant pH range. As will be shown below, the both monotonic and physically reasonable parametrization of $A_1$, $\tau_1$, $A_2$, and $\tau_2$ over a large range of reaction conditions in itself shows that the data are well-described by two parallel and exponentially decaying pathways.

Figure 2 and Supplementary Table 5 show how the bi-exponential decay depends on pH and H/D isotope exchange at a constant ionic strength (0.2 M) and 4% surface excitation; Figure 2A and 2B show the raw data and the fits. The relative amplitude ($\frac{A_n}{A_1 + A_2}$) (Figure 3C) of the slower route increases with decreasing pH, which means that the slower route is favored at lower pH. Over pH 7 to pH 13, the two time constants ($\tau_1 \sim$ 8 µs and $\tau_2 \sim$ 60 µs) are conserved (Figure 2D). The systematic trends with pH demonstrate that the decay is not related to charge transfer to anions in solution, which are varied to keep the ionic strength constant over the pH range (Figure 3). Past pH 13, both time constants exhibit a marked decrease, reflecting a faster mechanism, in accordance with many previous electrochemical investigations of $O_2$ evolution[37-39]. Further, a H/D kinetic isotope effect (KIE) demarcates a separate mechanism for the slower route. With deuterated water, the slower decay (Figure 2B) reflects a constant H/D KIE of ~1.75 on $\tau_2$ (Figure 2D). While a few H/D KIE's have been reported for water oxidation at electrode surfaces, they are based on time-resolved photo-current[38] or steady state product evolution[39] rather than the decay of an intermediate and do not exhibit the systematics seen here with reaction conditions.

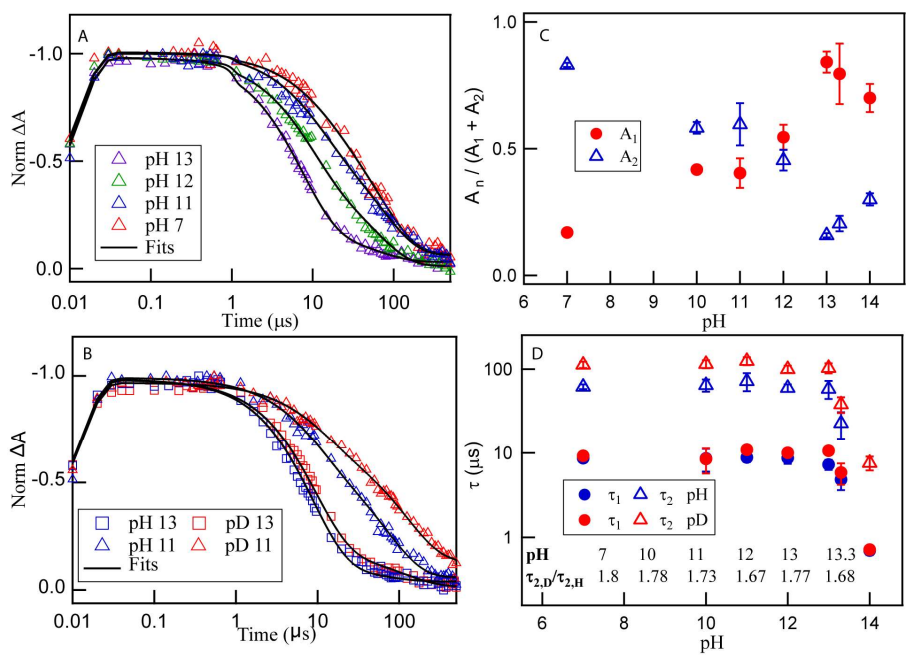

**Figure 2. pH dependence and H/D kinetic isotope effect (KIE) at 0.2 M ionic strength.** Normalized kinetics of the transient response at 0 V vs Ag/AgCl with a 400 nm probe fit with bi-exponential decays: (A) for different pH and (B) for pH/pD 11 and pH/pD 13. (C) pH-dependent relative amplitudes $[A_n/(A_1 + A_2)]$ of the two decay routes (D) pH-dependent time constants of the two decay routes ($\tau_1$, $\tau_2$) in both normal and deuterated water. The pronounced H/D KIE ~ 1.75 of the slower decay route ($\tau_2$) is denoted for each pH. Deuterated water does not change (within 6.5%) the amplitudes ($A_1$, $A_2$ for oxyl & bridge intermediates) associated with each decay route. In order to maintain a 0.2 M ionic strength of the electrolyte throughout the pH range, different buffers were utilized (pH 7, 12, 13 solutions buffered with $Na_2SO_4$, pH 10 and 11 buffered with $NaHCO_3$ and $Na_2CO_3$).

Figure 3 shows how the bi-exponential decay depends on ionic strength and surface excitation at pH 7. At constant surface excitation (7 x $10^{13}$ cm$^{-2}$ or 3% of O-sites), ionic strength dramatically tunes which route is favored (Figure 3A, 3B), with the slower route dominating at 0.02 M and with equal contributions of both by 1 M. Importantly, ionic strength does not modify the overall signal amplitude: the faster route's contribution to the decay increases directly at the expense of the slower route, conclusively identifying two parallel pathways (Figure 3B). An unchanged signal amplitude over a molarity range from 0 to 4 M also means that, while ionic strength can modulate the protonation of the surface[40], it is not significant enough to modify the initial conditions giving rise to the electronic levels of $A_1 + A_2$. We note that, in contrast, bulk pH does change the initial optical cross-section significantly, leading to the relative magnitudes ($\frac{A_n}{A_1 + A_2}$) reported. Therefore, a similar surface population of localized oxyl & bridge intermediates, created at picosecond times scales and determined by a given pH, exhibits a nanosecond branching into the two populations ($A_1$ and $A_2$).

Increasing ionic strength also decreases $\tau_1$ and $\tau_2$ and the effect saturates at 1 M (Figure 3C). Since the salt's ion density clearly tunes the reaction pathways, screening should modulate the activation barriers through which the initial population decays. In order to quantitatively account for screening, the ionic strength is plotted as the cation surface density of a sphere of the bulk electrolyte. For the electrolyte $Na_2SO_4$, the cation surface density, $[Na^+]_s$, is equivalent to the negative charge contained in the anion ($SO_4^{2-}$) surface density. Plotted this way, 1 M translates to $7 \times 10^{13}$ cm$^{-2}$ $[Na^+]_s$, which means that the ionic strength effects saturate when the anion charge density equals the surface hole excitation (Figure 3B, 3C). If the surface excitation decreases by half, the saturation point decreases accordingly (black vs. red in Figure 3B, 3C). Since the ionic strength was modulated at a constant and neutral pH, the anions ($SO_4^{2-}$) screen the population of one-electron intermediates at O-sites. These results implicate screening, in the planar sense, to the bifurcation of the population between two reaction pathways and the saturation of the rate constants.

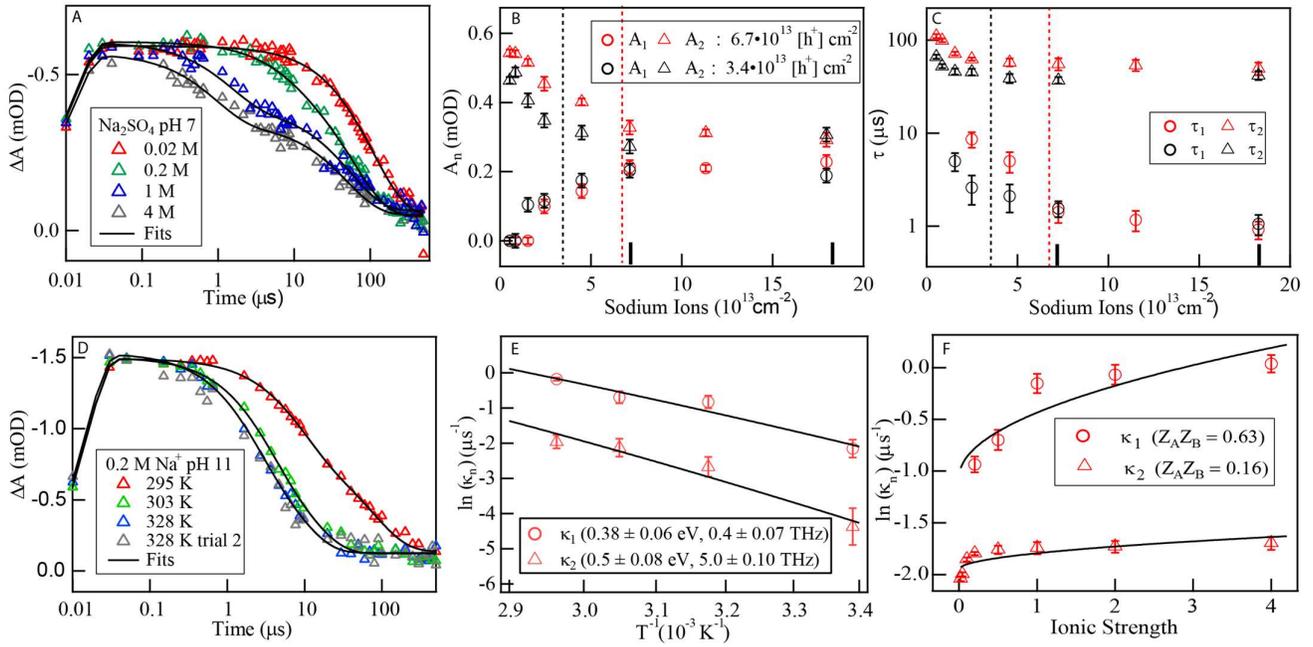

**Figure 3 Ionic strength and temperature dependence at constant pH.** (A) Kinetics of the transient response measured at pH 7 for increasing molarity of $Na_2SO_4$ and fit with bi-exponential decays. (B) The amplitudes of the two decay routes ($A_1$, $A_2$) plotted against the surface sodium ion density, $[Na^+]_s$, or equivalently the total negative surface charge coming from $SO_4^{2-}$ (the x-axis bars indicate 1 and 4 M $Na_2SO_4$). The amplitudes are shown for two surface hole excitations ($[h^+]$ $cm^{-2}$) (red, black). (C) The lifetime of the two decay routes ($\tau_1$, $\tau_2$) plotted against $[Na^+]_s$ for the two surface hole excitations (red, black). The vertical dotted lines in B, and C, indicate where $[Na^+]_s$ equals the surface hole density. Uncertainties for B and C come from the standard deviation associated with the fitting parameters. (D) Selected kinetics of transient response for the 0.1% Nb-doped $SrTiO_3$ in pH 11 solution with 0.2 M sodium ion concentration (buffered with $Na_2CO_3$ and $Na_2SO_4$) at 295K (red), 303K (green) and 328K (blue) with fits. Reproducible faster decay dynamics observed for increasing temperature as indicated by trial 2 at 328 K (grey). (E) Logarithmic (ln) dependence of decay rates ($\kappa_1$ ($\mu s^{-1}$) defined as ($1/\tau_1$) and $\kappa_2$ ($\mu s^{-1}$) defined as ($1/\tau_2$)) on 1/T. The slopes and intercepts of the fit (black line) to the data (red symbols) reveal the energies and pre-factors respectively. Error bars on the graph come from uncertainties in the fitted time constant after at least six different trials at the same temperature. The uncertainty of activation barriers and prefactors came from least square linear fitting. (F) Logarithmic (log) dependence of the decay rates ($\kappa_1$, $\kappa_2$) on the square root of the ionic strength, I, shown through the fit (black line) to the data (red symbols). The steepness of the dependence is defined by the charge on the reactants ($Z_AZ_B$) in the transition state.

Two decay rates systematically tuned by pH, ionic strength, and H/D exchange demonstrate reaction-dependent activation barriers. Transition state theory (TST) explicitly connects rate constants to a transition state (TS) by modulating reaction conditions that affect only the activation barrier, but not the population of reacting species.[41] In so doing, the theory allows a connection to the TS to be made without postulating a specific mechanism. Here, we explicitly explore first temperature and then ionic strength in this context. In TST, reaction rates ($\kappa_n$) are equal to $Ae^{-E_a/kT}$ with $E_a$ the activation energy, $A$ the reaction pre-factor, $T$ the temperature, $k$ Boltzman's constant, and $h$ Planck's constant. Figure 3D shows the decay dynamics for different temperatures at pH = 11 with 0.2 M sodium ion concentration (buffered with $Na_2CO_3$ and $Na_2SO_4$) at 0 V vs. Ag/AgCl. A pH 11 solution is chosen because at this pH the $\tau_1$ and $\tau_2$ decay pathways contribute fairly equally through $A_1$ and $A_2$, making it relatively easy to see the effect of temperature on both decay pathways. The dynamics were reproducible for 295K (room temperature) < T < 330 K as indicated by the 328 K curve trial 2 in Figure 3D. For temperature T < 280K, water condenses on the cell window which results in too much scattering of the probe light. For temperature T > 330 K, the time-traces were not as repeatable, suggesting that irreversible reactions at the sample surface changed the reaction kinetics. The decay dynamics were fitted consistently with the bi-exponential methods described in the SI to extract the reaction rate constants. The fitted reaction rates $\kappa_n$ ($1/\tau_n$) are shown in Figure 3E. With increasing temperature, both $\kappa_1$ and $\kappa_2$ increase, showing that the kinetics indeed increase with temperature as expected. More importantly, $\kappa_1$ and $\kappa_2$ are logarithmically (ln) dependent on $1/T$. The processes defining $\kappa_1$ are indeed different than $\kappa_2$, with $\kappa_1$ much larger than $\kappa_2$ for all temperature points on the logarithmic plot. Within TST, the difference either originates from the pre-factor $A$ or the activation barrier $E_a$. In a free linear fit of the data, that gives us the best $R^2$ (0.95), the activation barriers are $0.38 \pm 0.06$ eV and $0.5 \pm 0.08$ eV and the pre-factors are $0.4 \pm 0.1$ THz and $5 \pm 0.1$ THz respectively for $\kappa_1$ and $\kappa_2$. Within this free fit, the difference in kinetics arises from the lower activation barrier for $\kappa_1$. Since the activation barriers overlap within the error, we also performed the linear fit by fixing the activation barrier to the region of overlap (0.42-0.44 eV). With a fixed activation barrier for both routes, the pre-factor $A$ instead differentiates the faster from the slower route, with values of $4.9 \pm 0.1$ THz and $4.3 \pm 0.1$ THz respectively albeit with lower $R^2$ values (See Supplementary Figure 14 and 15 for detailed analysis of both fitting methods). The analysis clearly separates the two processes but cannot fully define the origin of the kinetic difference within the Arrhenius model. However, in all cases the slopes define activation energies of ~ 0.4-0.5 eV, consistent with that expected for O-O bond formation events by DFT calculations[42,43]. Further, the pre-factors are within 0.4 – 5 THz, which is the relevant range for TST that anticipates $\frac{kT}{h}$ ~6 THz for events where each relevant vibration transmits a reactant through the barrier.

If the reactants that create the transition state involve ions or trapped charge, the solution ionic strength can modulate the activation barrier by screening the charge distribution in the transition state. In order for screening to affect the barrier, at least two reactants must be involved. The Bronsted-Bjerrum equation[44,45] defines this primary kinetic salt effect and is given, for a bimolecular reaction, by:

$$\log \kappa = \log \kappa_0 + 1.02\, Z_A Z_B\, I^{1/2},$$

where $\kappa$ is the measured rate, $\kappa_0$ the rate in the absence of a salt, $Z_A$ and $Z_B$ the charge associated with reactants A and B, and I=ionic strength. $I = 0.5 \sum C_i Z_i^2$, where $C_i$ and $Z_i$ are the concentration of the solution ions and the charge on each respectively. Figure 3F re-casts the $\tau_1$, $\tau_2$ dependence on $[Na^+]_s$ on I, revealing a square-root relationship between the reaction rate and I and a positive $Z_A Z_B$. The positive $Z_A Z_B$ means that two reactants of the same type of charge are associated with the transition state, such that counter-ions screen them from each other and increase the rate through the barrier. Therefore, the Bronsted-Bjerrum equation confirms and furthers the above interpretation: solution anions screen interactions between two or more one-electron intermediates. Moreover, it connects this screening to the transition state. Using the bimolecular description as an example, the screening is stronger for the $\kappa_1$ ($Z_A Z_B = 0.63 \pm 0.18$, $\chi^2 = 0.15$) than the $\kappa_2$ ($Z_A Z_B = 0.16 \pm 0.04$, $\chi^2 = 0.04$) TS. $Z_A Z_B$ is less than 1 in both, which means that the charge on the O-sites is less than a full VB hole, as anticipated for charge trapping associated with proton transfers.

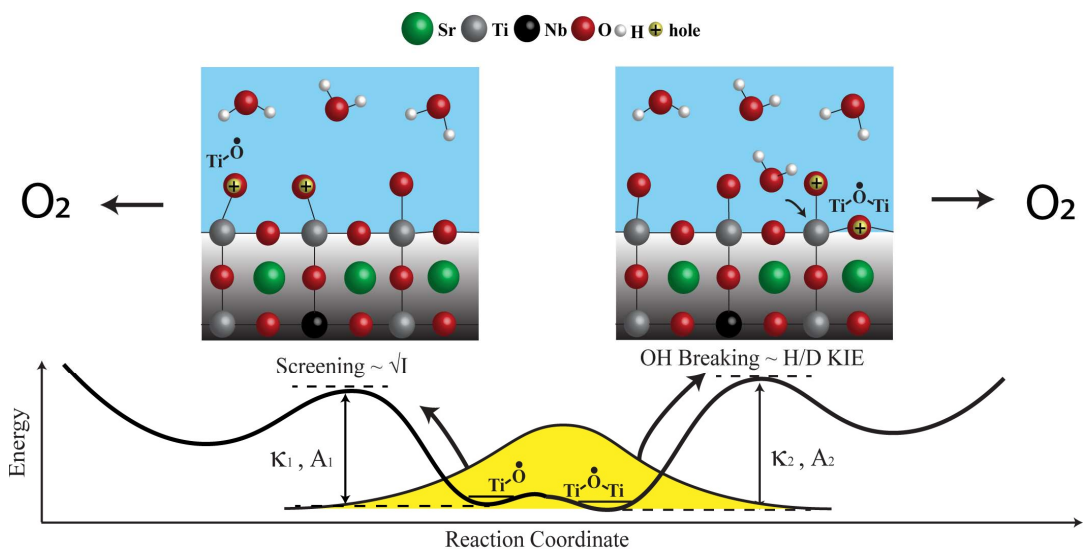

**Figure 4 Isolating two activation barriers** Cartoon and free energy description of the two reaction pathways described experimentally, with the distributed configurations of the principal intermediates represented as the yellow Gaussian distribution. The cartoon represents proposed two transition states towards $O_2$ evolution.

**Discussion**

The cumulative observations demonstrate TST dynamics of two local transition states of water oxidation at an electrode surface, which uniquely isolates a reaction barrier from the four expected within the full cycle. The transition states are inherently local because they are defined by two distinct reaction rates rather than many, one of which exhibits a unique H/D KIE, and both of which reflect screened transition states through the primary kinetic salt effect. That both routes are relevant for the water oxidation cycle is understood through the pathway dependence on reaction conditions, in the context of the experimental setup in which the initial charge-trapping intermediates achieve Faradaic $O_2$ evolution. The reaction conditions monotonically tune the pathways such that either can dominantly contribute to the decay associated with Faradaic $O_2$ evolution. This is especially clear with the dependence on ionic strength, which bi-furcates the initial population into the two pathways.

We now turn to the assignment of the two competing transition states, given one should relate to the decay of an oxyl site and the other, to a bridge site (Figure 4). The pathway ($A_1$, $A_2$) and rate ($\kappa_1$ ($1/\tau_1$), $\kappa_2$ ($1/\tau_2$)) dependence on pH and I assigns the faster $\kappa_1$ to the oxyl, since: (1) higher pH favors a dissociated Ti-$O^-$ terminated surface and so a higher relative population of oxyls (2) screening should affect primarily the labile oxyl radical (3) at very low ionic strengths, oxyl radicals are not screened and therefore the route of bond formation through them is not supported ($A_1 \sim 0$). All three observations are relevant to the commonly cited bi-radical recombination mechanism of O-O bond formation from two neighboring oxyl radicals[18,46]. Furthermore, this mechanism requires two or more localized intermediates in the TS, which is reflected in the primary kinetic salt effect. While observed for both pathways, the Bronsted-Bjerrjum equation has a larger $Z_A Z_B$ for the faster $\kappa_1$ pathway. On the other hand, the H/D KIE on $\kappa_2$ should occur with a distinct bond-breaking O-H event; this event is required in nucleophilic attack mechanisms between a lattice (bridge) oxygen and either $H_2O$ or $OH^-$[46-49]. While the value of ~1.75 is lower than that expected for a pure O-H breaking event and could be masked by screening, a similar H/D KIE has been measured for $O_2$ evolution from the water oxidation complex in Photosystem II[13,18]. We note that neither pathway exhibits a measurable $O^{16}/O^{18}$ KIE (Supplementary Figure 16), which is consistent with mechanisms that do not involve breaking an O-O bond.

We now turn to how these TST dynamics were revealed at an electrode surface. TST dynamics should reflect the full range of reactant configurations involved: in the reactant quasi-equilibrium description of TST, the transition state is in a quasi-equilibrium with the entire reactant population, which includes that site undergoing the transformation and how it interacts with neighboring sites. Therefore, the reactant population includes site-localized bridge and oxyl intermediates, from which an O-O bond might form, and their time-dependent configurations. Molecular dynamics simulations identify that diffusive surface-hopping should exist on titania surfaces: local exchange of a hole at neighboring O-sites turns a bridge intermediate into an oxyl, and vice-versa, on the picosecond time scale[24]. Such diffusive hopping should lead to many different configurations of oxyl and bridge intermediates. This is especially the case for 3% surface excitation, which separates hole-trapping intermediates from each other by at most ~2 unit cells, such that even short range hopping could lead to a large distribution. In this context, a broad band (~ 1 eV) optical probe of TST, which averages over all possible intermediate configurations through their mid-gap electronic levels, can detect the TST dynamics while the infrared probe of the normal modes of a specific site-localized intermediate does not. Indeed, the results advance that transition states exist separately from a specific reactant, since reaction conditions branch the populations into two separate population pathways, implying that the decay of any one site depends highly on the configuration of its neighboring ones. A very promising avenue for future work is to investigate this dichotomy, perhaps through the intermediates' surface hopping, in which the surface as a whole rather than a particular surface site leads to catalytic events. The distributed reactant configurations that can reach either transition states are depicted in yellow in Figure 4.

In summary, we demonstrate competing pathways by which the charge-trapping intermediates of the water oxidation reaction on titania surfaces, the oxyl and bridge, decay towards the next step in the reaction using canonical transition state theory. The competing pathways reflect what has often been thought of as important for electrode surfaces: the multitude of sites allows for multiple mechanisms. One of the challenges has been to identify the transition states clearly enough to understand how to tune between disparate mechanisms. By isolating the intermediates' decay, we demonstrate how to use reaction conditions (pH, ionic strength) to select between two competing transition states. The ionic strength is especially interesting, since it can branch the same initial population of oxyl & bridge intermediates fully towards the slower route or create equal populations through the fast and slow routes.

**Methods:**

**Samples**

0.08% Nb-doped $SrTiO_3$ by weight (henceforth 0.08% is referred to as 0.1%) $SrTiO_3$ single crystals with crystal-lographic orientation (100) were obtained from MTI Corp. (Richmond, CA). The crystals were 0.5 mm thick with polished front sides (Ra < 5 Å) and unpolished backsides. All spectroscopic measurements were performed on the polished front sides. All spectroelectro-chemical measurements were performed in a Teflon electrochemical cell with $CaF_2$ optical windows (3 mm thick). The electrolyte was in contact with the atmosphere during measurements. The potential of the n-$SrTiO_3$ photoelectrode with respect to an Ag/AgCl (3M KCl) reference electrode (MF-2052; Basi, West Lafayette, IN) was controlled by a CHI650E Potentiostat (CH Instruments, Austin, TX). A Pt wire served as the counter electrode. Ohmic contact between the unpolished n-SrTiO$_3$ backside and copper wire was established using Gallium-Indium eutectic (Sigma-Aldrich, St. Louis, MO). For the transient reflectance experiments, an insulating lacquer covered all surfaces except the polished front side of the crystal. The exposed front surface areas of the 0.1% and Nb-doped samples was 25 mm$^2$.

**Transient experiments**

For the transient optical experiments, the pump beam was derived from a Nd:YAG laser system producing pulses with a center wavelength of 1064 nm and ≈ 40 ns temporal width at a 1 kHz repetition rate. All the output was directed into a Fourth harmonic generation setup to generate 266 nm light as pump. The probe beam was derived from a regeneratively amplified Ti: sapphire laser system (Coherent Legend; Coherent, Inc., Santa Clara, CA) producing pulses with a center wavelength of 800 nm and ≈ 150 fs temporal width at a 1 kHz repetition rate. Part of the 800 nm beam was frequency doubled to generate 400 nm light or focused into a sapphire crystal to generate white light as probe. The detailed sample configuration could be seen in Supplementary Figure 1. The polarization of the 400 nm probe beam was controlled by changing the 800 nm polarization using a ½-wave plate and a linear polarizer. The pump beam was incident normal to the sample surface. After the sample, the reflected probe beam was focused into an optical fiber, which was coupled to a CMOS array spectrometer (CAM-VIS-3; Ultrafast Systems, LLC, Sarasota, FL). In all experiments, the pump beam was modulated by a mechanical chopper (3501; Newport, Inc., Ir-vine, CA) at a frequency of 500 Hz. The detector output was interfaced with a personal computer, which provided automated control over an electronic delay generator (DG645; Stanford Research Systems, Sunnyvale, CA). The typical incident pump fluence was about 0.05 mJ cm$^{-2}$ at 266 nm, corresponding to $6.7 \times 10^{13}$ photons cm$^{-2}$. For a typical n-SrTiO$_3$ (100) surface, with 3.905 Å as lattice constant, the number of oxygen atoms on the surface is close to $1.97 \times 10^{15}$ atoms cm$^{-2}$. Therefore, for a 0.05 mJ cm$^{-2}$ fluence at 70% quantum efficiency for charge-separation, the surface excitation is close to 3.4% (counting oxygen atoms). The excitation beam spot size was ~500 μm (fwhm).

Transient spectroscopy measurement was conducted in an electrochemical cell in reflection geometry. Six scans were performed and averaged to extract a single kinetic trace. Each scan was performed on a fresh sample spot and took less than 90 s to avoid any sample surface roughening.[30] The temperature of the electrochemical cell was controlled by a heating gun. Solution temperature was monitored with a thermal couple before and after the experiment to make sure the temperature was stable. In general, the temperature was stable within ± 3 K°. For transient infrared measurement, the pump beam was derived the same way as the optical measurement. Other procedures were kept the same with previous infrared measurements[20].

**Data Availability:** Representative data and all of the analysis from the extended data set supporting the findings are within the paper (and its supplementary files). The extended data set supporting the findings is available from the corresponding author upon reasonable request.

**Acknowledgments**: The experimental work was supported by the Director, Office of Science, Office of Basic Energy Sciences, and by the Division of Chemical Sciences, Geosciences and Biosciences of the U.S. Department of Energy at LBNL under Contract No. DE-AC02-05CH11231. We thank Heinz Frei and Joel Eaves for fruitful discussions.

**Competing Interests** The authors declare that they have no competing financial interests.

**Contributions** T.C. and X.C. conceived the project and T.C. wrote the manuscript with input from all authors. X.C and D.A constructed the transient setup, collected transient data and prepared samples. X.C and T.C. analyzed the transient data.

**Correspondence**     Correspondence and requests for materials should be addressed to Tanja Cuk (email: tanja.cuk@colorado.edu).

**Supplementary Information**     Supplementary information is available in the online version of the paper.